\documentclass[12pt,epsf]{article}
\usepackage{graphicx, amsmath}

\begin{document}
\sloppy
\begin{flushright}{SIT-HEP/TM-53}
\end{flushright}
\vskip 1.5 truecm
\centerline{\large{\bf Evolution of the curvature perturbations during
warm inflation}}
\vskip .75 truecm
\centerline{\bf Tomohiro Matsuda\footnote{matsuda@sit.ac.jp}}
\vskip .4 truecm
\centerline {\it Laboratory of Physics, Saitama Institute of Technology,}
\centerline {\it Fusaiji, Okabe-machi, Saitama 369-0293, 
Japan}
\vskip 1. truecm

\makeatletter
\@addtoreset{equation}{section}
\def\theequation{\thesection.\arabic{equation}}
\makeatother
\vskip 1. truecm

\begin{abstract}
\hspace*{\parindent}
This paper considers warm inflation as an interesting application of  
multi-field inflation. Delta-N formalism is used for the calculation of
 the evolution of the curvature perturbations during warm
inflation. Although the perturbations considered in this paper are
decaying after the horizon exit, the corrections to the curvature
 perturbations sourced by these
perturbations can remain and dominate the curvature perturbations at
large scales. In addition to the typical evolution of the curvature
perturbations, inhomogeneous diffusion rate is considered for warm
inflation, which may lead to significant non-Gaussianity of the
spectrum.
\end{abstract}

\newpage
\section{Introduction}
Inflation is the most successful model for the large
scale structure of the Universe in terms of the very early Universe.
An inflationary Universe is consistent with current observations of the
temperature anisotropy of the cosmic microwave background (CMB).
Inflationary dynamics may be categorized in two ways: the original
(supercooled) inflation scenario, and warm inflation
\cite{warm-inflation-original}. 
In the warm inflation scenario, dissipative effects are significant
and radiation production occurs concurrently during the inflationary
period\cite{Hosoya-Sakagami}. 
Basically, the spectrum of the cosmological perturbations created during
inflation is expected to be scale-invariant and Gaussian.
However, there are anomalies in the spectrum, such as a small  
departure from exact scale-invariance and a certain non-Gaussian
character \cite{Bartolo-text, NG-obs}, both of which can help
reveal the underlying theory and dynamics of the fields during
 inflation. 
The observation of a small shift in the spectrum
index $n-1\ne 0$ is an obvious example\cite{EU-book} of
scale invariance violation.
Observations may also support a non-Gaussian character in the spectrum.
Determining how these anomalies arise and how they relate to
the inflation dynamics and character of the underlying theory, 
will depend on the model used.
There are many models for inflation, in which the spectrum is generated
(1) during inflation \cite{Modulated-matsuda, Kofman-modulated, A-NEW,
matsuda-stop-index, roulette-inflation},
(2) at the end of 
inflation \cite{End-Modulated, End-multi, End-multi-mat},
  (3) after inflation by preheating\cite{IH-PR, kin-NG-matsuda, 
IH-string} and reheating\cite{IH-R, Preheat-ng}    
 or (4) late after inflation by curvatons \cite{curvaton-paper,
 curvaton-dynamics, Mazumdar-curvaton, matsuda_curvaton}
or by inhomogeneous phase transition\cite{IH-pt}. 

For warm inflation, correction arises during evolution of the curvature
perturbations that may be a significant source of these anomalies.
We thus study the evolution of the curvature perturbations during warm
inflation.
In warm inflation, the effective potential of
the inflaton fields depends on temperature $T$, which can be expressed
as $V(\phi_i,T)$. 
The diffusion rate $\Gamma_i$ may also depend on $T$ and $\phi_i$.
Therefore, the ``trajectory'' of warm inflation
is given by both inflaton fields $\phi_i$ and $T$.
The situation reminds us of multi-field inflation, in which
the dynamics of a inflaton field is determined by the 
      other inflaton fields, and vice versa.\footnote{See also
      Ref.\cite{simulation} in which the
      evolution of the curvature perturbations is studied using
 numerical calculation.}
The mutual dependency is important in the analysis of multi-field
inflation.
For warm inflation, the mutual dependency between inflaton fields
and $T$ is important.
This paper considers warm inflation as an interesting application of 
 multi-field inflation. 
The evolution of linear perturbations about a FRW background 
metric has been discussed for supercooled inflation 
in Ref.\cite{A-NEW}, using only the
      local conservation of energy-momentum. 
The result does not depend on details of the inflation dynamics,
thus it can be applied for warm inflation.
In this paper, we apply the $\delta N$ formula\cite{A-NEW} 
to warm inflation. 
The definitions of the $\delta N$ formula are discussed
in the appendix of this paper. 

We find that there is a significant correction to the curvature
perturbation {\bf if the curvature perturbation at horizon crossing 
is expressed by the standard gauge-invariant formula.} 
We also find that a significant correction may arise from the
perturbations of the radiation  density.

\underline{Evolution of curvature perturbation in the $\delta N$
formalism}

Evolution of the curvature perturbation can be studied without 
reference
to specific inflationary dynamics or gravitational field
equations\cite{A-NEW}. 
In the following we demonstrate the calculation of the evolution of the
curvature perturbations using the $\delta N$ formalism\cite{Modulated-matsuda}.
Calculations under this method are quite straightforward compared with
calculations based on the time-derivative of the curvature perturbations
on spatially flat hypersurfaces.

In standard (supercooled) inflation, the spectrum of the curvature
perturbation ${\cal P}_{\cal R}(k)$ for the (adiabatic) inflaton field
$\phi$ is given by 
\begin{equation}
\label{bound-ini}
{\cal P}_{\cal R}(k) = \left(\frac{H}{\dot{\phi}}\right)^2
\left(\frac{H}{2\pi}\right)^2,
\end{equation}
where the right-hand side is evaluated at the epoch of the horizon exit
$k=aH$.
Here $H$ is the Hubble parameter and $a$ is the cosmic-scale factor.
Here the comoving curvature perturbation ${\cal R}$ is considered, 
which is related to the curvature perturbation on uniform-density
hypersurfaces $\zeta$ as ${\cal R}\simeq -\zeta$ at large
scales. 
The above equation is not exact in the warm inflation scenario, if 
the gauge-invariant combinations for the curvature
perturbation are defined by
\begin{eqnarray}
\label{zeta-org}
\zeta &=&-\psi -H\frac{\delta \rho}{\dot{\rho}}\nonumber\\
{\cal R}&=& \psi -H\frac{\delta q}{\rho+p},
\end{eqnarray}
where $\delta q =-\dot{\phi}\delta
\phi$  is the momentum perturbation satisfying
\begin{equation}
\epsilon_m=\delta \rho-3H \delta q.
\end{equation}
Here $\epsilon_m$ is the perturbation of the comoving density,
satisfying the evolution equation 
\begin{equation}
\label{decay-comv}
\epsilon_m =-\frac{1}{4\pi G}\frac{k^2}{a^2}\Psi,
\end{equation}
where $\Psi$ is related to the shear perturbation and assumed to be
finite in this paper.
The $\delta q=0$ hypersurfaces are identical to uniform density 
hypersurfaces ($\delta \rho=0$) 
at large scales ($\epsilon_m\equiv \delta \rho-3H\delta q\simeq 0$).
Linear scalar perturbations of a
Friedman-Robertson-Walker(FRW) background are considered:
\begin{equation}
ds^2=-(1+2A)dt^2 + 2a^2(t)\nabla_i B dx^i dt +a^2(t)
[(1-2\psi)\gamma_{ij}+2\nabla_i\nabla_j E]dx^i dx^j.
\end{equation}
Here $\rho$ and $p$ denote the energy density and the pressure 
during inflation.
The comoving curvature perturbation on spatially flat hypersurfaces
$\psi=0$ is expressed as 
\begin{eqnarray}
{\cal R}&=&  -H\frac{\delta q}{\rho+p}\nonumber\\
&=&  -H\frac{\delta q}{\dot{\phi}^2 + Ts}\nonumber\\
&\sim &  -H\frac{\delta q}{\dot{\phi}^2}\times (1+r_\Gamma)^{-1},
\end{eqnarray}
where the differences from the standard supercooled inflation
arise in the entropy ($s$) and the damping term ($\Gamma$), as will be
defined and discussed in Sec.2.
The difference between the one obtained in previous study of warm
inflation will be
 explained in Sec.2 by considering the evolution of the curvature
 perturbations. 
In warm inflation, the inflation trajectory is determined by both 
$\phi$ and $T$, which leads to an interesting realization of
``multi-field inflation''.
In fact, in supercooled inflation, multi-field model is 
considered using the total energy density.
In this respect, it would be interesting to consider warm inflation
as an interesting application of multi-field inflation.

The evolution of the curvature perturbation at large scales
is found to be given by the non-adiabatic pressure perturbation
$\delta p_{nad}$; 
\begin{equation}
\label{111}
\dot{\zeta}\simeq -H\frac{\delta p_{nad}}{\rho+p},
\end{equation}
where $\zeta$ and ${\cal R}$ coincide ($\zeta\simeq-{\cal R}$)
at large scales and 
$\delta p_{nad}\equiv\left[\delta p
-\frac{\dot{p}}{\dot{\rho}}\delta \rho\right]$
is a gauge-invariant perturbation.
This equation is valid independent of gravitational field equations.
In fact, it can be found from Eq.(\ref{zeta-org}) that
\begin{equation}
\label{t-der-zeta}
\dot{\zeta} =-\dot{\psi}-\frac{d}{dt}\left[
H\frac{\delta \rho}{\dot{\rho}}\right].
\end{equation}
Equations for the local conservation of energy momentum
lead to the following useful expansion: 
\begin{equation}
\label{enemoeq}
\dot{\delta \rho}=-3H(\delta \rho+\delta p)
+(\rho+p)\left[3\dot{\psi}-\nabla^2 (\sigma + v + B)\right],
\end{equation}
where the scalar describing the shear is
\begin{equation}
\sigma = \dot{E}-B
\end{equation}
and $\nabla ^i v$ is the perturbed 3-velocity of the fluid.
Eq. (\ref{enemoeq}) gives the equation for $\dot{\psi}$ in terms of the
local conservation of energy momentum, which gives the simple
equation for the evolution of $\zeta$:
\begin{eqnarray}
\label{t-der-zeta2}
\dot{\zeta} &\simeq&
-\frac{\dot{\delta \rho}+3H(\delta \rho + \delta p)}{3(\rho+p)}
+\frac{d}{dt}\left[H\frac{\delta \rho}{3H(\rho+p)}\right]\nonumber\\
&=& -\frac{H}{\rho+p}
\left[\delta p-\frac{\dot{p}}{\dot{\rho}}\delta \rho \right]\nonumber\\
&=& -\frac{H}{\rho+p} \delta p_{nad} 
\end{eqnarray}
where $\nabla^2 (\sigma + v + B)$ is
neglected.
To compare the above result with the $\delta N$
formalism\cite{delta-N-ini}, it is useful to define the
perturbed expansion rate with respect to the 
coordinate time
\begin{equation}
\delta \tilde{\theta}\equiv -3\dot{\psi}+\nabla^2\sigma,
\end{equation}
which can be used to define the $\delta N$ formalism.
Choosing a gauge with a flat slicing at $t_{ini}$ and uniform
density at 
$t$, and using $\zeta =-\psi$ for the specific choice of slice at
$t$, the $\delta N$ formula is given by
\begin{equation}
\zeta = \frac{1}{3}\int^t_{t_{ini}}\delta \tilde{\theta}dt =\delta N.
\end{equation}
The evolution equation for the $\delta N$ formalism is given by
\begin{eqnarray}
\label{del-n-t-2}
\frac{d}{dt}
\delta N &\equiv& \frac{1}{3}\delta \tilde\theta \nonumber\\
&\simeq& -\dot{\psi}\nonumber\\
&=& \dot{\zeta} +
\frac{d}{dt}\left(H\frac{\delta \rho}{\dot{\rho}}\right)
\nonumber\\
&=&-\frac{H}{\rho+p} \delta p_{nad} +
\frac{d}{dt}\left(H\frac{\delta \rho}{\dot{\rho}}\right),
\end{eqnarray}
which shows that Eqs. (\ref{t-der-zeta2}) and (\ref{del-n-t-2}) are
consistent, as far as the relation is defined for the evolution of the 
curvature perturbation $\dot{\zeta}$ on a uniform density slice at $t$.
The equation for the perturbed expansion rate $\delta
\tilde{\theta}$ for $\delta N$ is practically valid for any
gauge and slicing, but the $\delta N$ formula that explains the relation 
between $\zeta$ and $\delta N$ is defined for the specific choice of
slice at $t$. 
The difference in the definition of the hypersurfaces will also lead to
the difference in the boundary perturbation at the end of inflation, and
the total $\delta N$ does not depend on the choice of
hypersurfaces.\footnote{See Ref.\cite{gen-gra-mod} for an interesting
example.} 
In fact, the comoving density perturbation expressed in uniform density
gauge shows that 
\begin{equation}
\epsilon_m =-3H\delta q =-\frac{1}{4\pi G}\frac{k^2}{a^2}\Psi,
\end{equation}
which suggests that the perturbation of the adiabatic field $\delta
\phi$ decays on uniform-density hypersurfaces.
This result is consistent with 
the standard inflationary scenario in which $\phi_e$ that defines the
the end of inflation is not perturbed.
The decaying $\delta \phi$ is not important in supercooled inflation,
 while it causes significant evolution of the curvature perturbation
in warm inflation, as will be discussed in Sec.2.
In the standard inflationary scenario, $\delta N$ is created by the
fluctuations of the inflaton field at the beginning of inflation.
For the analysis of warm inflation in terms of the $\delta N$ formalism, 
it is useful to define $\dot{\zeta}_{N}$ using Eq.(\ref{enemoeq})
as
\begin{eqnarray}
\label{orig}
\dot{\zeta}_{N} &\equiv& \frac{d}{dt}\delta N\simeq -\dot{\psi}
\nonumber\\
&\simeq&-
H\frac{\delta (\rho+p)}{(\rho+p)}-H\frac{\dot{\delta \rho}}{3(\rho+p)}
\nonumber\\
&=&-
H\frac{\delta (\rho+p)}{(\rho+p)}\nonumber\\
&=& -H\frac{2\delta K + \delta (Ts)}{2K + Ts},
\end{eqnarray}
where $Ts$ appears for warm inflation but does not appear for
supercooled inflation.
This equation will be discussed in Sec.2.

As a result, in terms of the $\delta N$ formalism, the evolution of the
curvature perturbation in warm inflation can be
 explained using the perturbations related to the 
inflaton kinetic term ($\delta K$) and the radiation density
$\delta(Ts)$, where the radiation density is given by $\rho_r=3Ts/4$.
In this paper, using the $\delta N$ formalism,
 a simple method for calculating the evolution of the curvature
 perturbation is considered for warm inflation.

\section{Evolution of the curvature perturbation}

\subsection{Warm inflation}

We first consider a homogeneous inflaton field interacting with
thermal radiation\cite{Hall-Moss-Berera}.
Here we mainly follow the arguments in Ref.\cite{Hall-Moss-Berera}.
The thermodynamic potential is given by
\begin{equation}
V(\phi, T)= -\frac{\pi^2}{90}g_*T^4 + \frac{m(\phi,T)\phi^2}{2}
+V_0(\phi),
\end{equation}
where $g_*$ is the effective number of thermal particles.
The evolution equation for the inflaton field is given by
\begin{equation}
\ddot{\phi}+(3H+\Gamma)\dot{\phi}+V(\phi,T)_\phi=0,
\end{equation}
where $\Gamma$ is the damping terms and $H$ is the expansion rate of the
Universe. 
Here the subscript denotes the derivative with respect to the field.
The strength of the thermal damping is measured by the rate $r$ given by
\begin{equation}
r_\Gamma\equiv \frac{\Gamma}{3H},
\end{equation}
which can be used to rewrite the field equation as
\begin{equation}
\ddot{\phi}+3H(1+r_\Gamma)\dot{\phi}+V(\phi,T)_\phi=0.
\end{equation}
The typical situation for warm inflaton is defined by $r_\Gamma\gg 1$.
The production of entropy during inflation is caused by the dissipation
of the inflaton motion, where the entropy $s$ is defined by the
thermodynamic equation;
\begin{equation}
s\equiv - V(\phi,T)_T,
\end{equation}
where the subscript denotes the derivative with respect to the
temperature.
From the energy-momentum tensor, it is found that the energy density
$\rho$ and the pressure $p$ are given by
\begin{eqnarray}
\rho &=& K + V + Ts\nonumber\\
p&=& K-V,
\end{eqnarray}
where $K\equiv\frac{1}{2}\dot{\phi}^2$ is the contribution from the 
kinetic term of the inflaton field, and
the Friedman equation follows
\begin{equation}
H^2 = \frac{1}{3M_p^2}\left(K+V+Ts\right).
\end{equation}
The slow-roll approximation is very useful in estimating the order of
magnitude of the physical quantities:
\begin{eqnarray}
\dot{\phi}&\simeq& -\frac{V_\phi}{3H(1+r_\Gamma)}\\
Ts&\simeq& 2 r_\Gamma K\\
\rho+P &=& 2K +Ts \simeq 2(1+r_\Gamma)K,
\end{eqnarray}
where the second equation is obtained from the evolution equation for the
radiation energy density.
On spatially flat hypersurfaces $\psi=0$, the
comoving curvature perturbation created during warm inflaton
 is expressed as
\begin{eqnarray}
\label{normal-curv}
{\cal R}&=&  -H\frac{\delta q}{\rho+p}\nonumber\\
&=&  -H\frac{\delta q}{\dot{\phi}^2 + Ts}\nonumber\\
&\sim &  -H\frac{\delta q}{\dot{\phi}^2}\times (1+r_\Gamma)^{-1}.
\end{eqnarray}

The important source of density perturbations in warm inflation is
thermal fluctuations.
A comoving mode of thermal fluctuations during warm
inflation is created by thermal noise.
The behavior of a scalar field interacting with radiation can be
studied by using the Langevin equation
\begin{equation}
-\nabla \phi(x,t)+\Gamma \dot{\phi}(x,t)+V_\phi=\xi(x,t),
\end{equation}
where $\xi$ denotes a source of stochastic noise.
From the Langevin equation, the root-mean square fluctuation amplitude
of the inflaton field $\delta \phi$ after the freeze out at 
$k/a\sim (\Gamma H)^{1/2}$ is obtained to be
\begin{equation}
\label{pert-phi}
\delta \phi\sim -\left(\frac{\pi}{4}\right)
(\Gamma H)^{1/4}T^{1/2}
\sim r_\Gamma^{1/4}r_T^{1/2}H,
\end{equation}
where $r_T$ denotes the ratio between $T$ and $H$, defined by
 $r_T\equiv T/H$.
Thermal fluctuation of the radiation density is also 
important in warm inflation.
It has been calculated to be\cite{Hall-Moss-Berera}
\begin{equation}
\delta \rho_r\sim g_*^{1/2}\left(\frac{k}{a}\right)^{3/2}T^{5/2}.
\end{equation}

In warm inflation with single inflaton field, the total energy density
and pressure perturbations 
for the first-order approximation are given by\cite{Hall-Moss-Berera}
\begin{eqnarray}
\delta \rho &\simeq&
 \delta K +V_\phi \delta \phi + T\delta s 
\nonumber\\
\delta p &\simeq& \delta K -V_\phi \delta \phi + s\delta T.
\end{eqnarray}
For two-field inflation with ($\phi_1, \phi_2$), 
one may redefine the adiabatic inflaton field
$\phi_\sigma$ and the entropy field $\phi_s$.
Then $V(\phi_\sigma,\phi_s,T)\simeq V_{\phi_\sigma} \delta {\phi_\sigma}
+V_{\phi_s} \delta {\phi_s} + V_T \delta T$ is obtained, which leads to
\begin{eqnarray}
\delta \rho &\simeq&
 \delta K +V_{\phi_\sigma} \delta {\phi_\sigma} +
V_{\phi_s} \delta {\phi_s} + T\delta s 
\nonumber\\
\delta p &\simeq& \delta K -V_{\phi_\sigma} \delta \phi 
-V_{\phi_s} \delta {\phi_s} + s\delta T,
\end{eqnarray}
where the entropy field satisfies $\dot{\phi}_s=0$.

Assuming that $\phi$ is identified with the adiabatic inflaton field,
the momentum perturbation is given by 
\begin{equation}
\delta q = -\dot{\phi}\delta \phi.
\end{equation}
Considering the backreaction from radiation, the velocity perturbation
is given by\cite{warm-inflation-original}
\begin{equation}
\delta \dot{\phi}\sim \frac{k^2}{a^2H^2} H^2 r_\Gamma^{-3/4}r_T^{1/2}.
\end{equation}
This may lead to second-order perturbation
$\delta^{(2)} q \simeq H^2 r_\Gamma^{-3/4}r_T^{1/2}
\delta \phi \times k^2/(aH)^2$.
The second-order perturbation $\delta q^{(2)}$ may also appear for fields
that are not inflaton.
The second-order perturbation  is disregarded in this paper for
simplicity, but it may become significant when multiplied
by a huge number of the fields in the model.
Note also that the momentum perturbation proportional to $k^2/a^2$ 
vanishes at large scales.
The comoving density perturbation is defined by a gauge-invariant quantity
\begin{eqnarray}
\epsilon_m &\equiv& \delta \rho-3H\delta q\nonumber\\
&=&\delta K + \left(\delta V +\delta[Ts] - (1+r_\Gamma)^{-1} 
V_\phi\delta \phi \right) -(1+r_\Gamma)^{-1} \ddot{\phi}\delta \phi,
\end{eqnarray}
where the last equation is obtained using the field equation.
The comoving density perturbation satisfies the evolution equation 
\begin{equation}
\epsilon_m =-\frac{1}{4\pi G}\frac{k^2}{a^2}\Psi,
\end{equation}
where $\Psi$ is related to the shear perturbation.
Here we assume $\Psi$ reaches a finite value at large scales, which leads
to $\epsilon_m\simeq 0$.
To estimate $\dot{\zeta}_N$, the perturbation $\delta K$ needs to be 
obtained from the above equations. 
It is given by
\begin{equation}
\delta K \simeq -\delta V-\delta[Ts]+(1+r_\Gamma)^{-1}V_\phi\delta \phi,
\end{equation}
where $\ddot{\phi}$ has been disregarded.
The perturbation $\delta K$  in the $\delta N$ formula, which is caused
by perturbations in the kinetic term, can thus be replaced 
using the equation for the comoving energy density perturbation.
It follows that
\begin{eqnarray}
-\dot{\psi} &\simeq&
H\frac{2 \left\{\delta V +\delta[Ts]- (1+r_\Gamma)^{-1} 
V_\phi\delta \phi \right\} - \delta (Ts)}{2(1+r_\Gamma)K}\nonumber\\
&\simeq&-H\frac{3s \delta T+T\delta s }{2(1+r_\Gamma)K}
+2H\frac{V_{\phi_s}\delta \phi_s}{(1+r_\Gamma)K}
+H\frac{r_\Gamma V_\phi}{(1+r_\Gamma)^2K}\delta \phi.
\end{eqnarray}
The second term is important when there is a bend in the inflation 
trajectory.\footnote{We considered two-field warm inflation to show what
happens at the bend of the trajectory in warm inflation.}
In the third term $\delta \phi$ is defined on uniform-density
hypersurfaces. 
Considering two different definitions of $\delta t$, which are motivated
 by the $\delta N$ formula defined for different hypersurfaces;
\begin{eqnarray}
\delta t_{\delta \phi=0}&\equiv& \frac{\delta \phi}{\dot{\phi}}\\
\delta t_{\delta \rho=0}&\equiv& \frac{\delta \rho}{\dot{\rho}}
\simeq -\frac{\delta q}{\rho+p}\simeq
\frac{\delta \phi}{(1+r_\Gamma)\dot{\phi}},
\end{eqnarray}
we find that there is a significant discrepancy between 
these two definitions.
Namely, for strongly dissipating warm inflation (SDWI, $r_\Gamma \gg
1$),  we find 
$\delta t_{\delta \phi=0} \gg \delta t_{\delta \rho=0}$ at large scales
where $\epsilon_m\simeq 0$.
This discrepancy cannot appear in supercooled inflation, in which the
diffusion rate vanishes by definition.
As a result, in supercooled inflation, there is no significant evolution
caused by $\delta \phi$, but in strongly dissipating warm inflation, 
$\delta \phi$ in the third term is still as large as
Eq. (\ref{pert-phi}) and raises significant contribution.
 
Using the above equations and the $\delta N$ formula, it is possible to
examine the order of magnitude of $\dot{\zeta}_N$ that can be 
caused by perturbations of the radiation and the third term.
They can be estimated as
\begin{eqnarray}
\label{ddzeta_n}
\dot{\zeta}_N^{(rad)}
&\sim& H\frac{\delta \rho_r}{(1+r_\Gamma) K}\nonumber\\
&\sim& H\left(\frac{k}{aH}\right)^{3/2}
\frac{T^{5/2}H^{3/2}}{T^4}\frac{r_\Gamma}{1+r_\Gamma}\nonumber\\
&\sim& H\left(\frac{k}{aH}\right)^{3/2}
\frac{r_\Gamma}{r_T^{3/2}(1+r_\Gamma)}
\end{eqnarray}
and
\begin{eqnarray}
\label{ddzeta_3rd}
\dot{\zeta}_N^{3rd}&\sim& 
H^2\frac{r_\Gamma V_\phi \delta \phi}{(1+r_\Gamma)^2 K}
\nonumber\\
&\sim& H^2
\frac{r_\Gamma}{1+r_\Gamma}\frac{\delta \phi}{\dot{\phi}}.
\end{eqnarray}
Although the correction arising from $\delta \rho_r$ decays at large
scales as $\delta \rho_r\propto (k/a)^{3/2}$, it may 
cause significant correction during warm inflation.
Namely, a decaying effect expressed by $\dot{\zeta}_N\propto
H\delta Ce^{-AHt}$ can leave a significant effect after a
time-integral;
\begin{equation}
\Delta \zeta_N(t_*) \sim \int_0^{t_*} H\delta Ce^{-AHt} dt
\sim \frac{C}{A}\left(1-e^{-AHt_*}\right).
\end{equation}
Then, due to the constancy of the curvature perturbations at large
scales, the correction $\Delta \zeta_N$ remains at large
scales. 
Applying the above result to Eq.(\ref{ddzeta_n}), 
$\Delta \zeta_N^{(rad)}$ at large scales is found to be given by
\begin{equation}
\Delta \zeta_N^{(rad)} \sim \int_0^{t} \dot{\zeta}_N^{(rad)} dt
\sim \frac{r_\Gamma}{r_T^{3/2}(1+r_\Gamma)},
\end{equation}
where $k=aH$ at $t=0$.
We also find for the third term;
\begin{equation}
\Delta \zeta_N^{(3rd)} \sim \int_0^{t} \dot{\zeta}_N^{(3rd)} dt
\sim \frac{r_\Gamma^{7/4}}{r_T^{3/2}(1+r_\Gamma)}.
\end{equation}
$\Delta \zeta_N^{(3rd)}$ dominates the curvature perturbations for
$r_\Gamma\gg 1$, 
but these two terms are comparable when $r_\Gamma \simeq 1$,
near the boundary between the weakly dissipating ($r_\Gamma\ll1$) and
strongly dissipating ($r_\Gamma \gg 1$) phases of warm inflation.

Considering the CMB normalization ${\cal P_R}^{1/2}\sim 10^{-5}$,
it is straightforward to find that $\Delta \zeta_N$ may dominate the
curvature perturbation for realistic parameter space for the warm
inflation scenario. 
To avoid the excessive creation of the curvature perturbation in terms
of the evolution after the horizon exit,
the condition $r_T^{3/2}(1+r_\Gamma)r_\Gamma^{-7/4}>10^5$ must be 
imposed for warm inflation.
For large $r_\Gamma$ and $r_T$, the result can be simplified to give
the condition $r_T^{3/2}r_\Gamma^{-3/4}>10^{5}$, which is required to satisfy
$\Delta \zeta_N<10^{-5}$ but leads to a significant reduction of the
conventional curvature perturbations created at the horizon exit.

There are two important consequences in the analysis.
One is that the {\bf conventional 
curvature perturbation} defined by ${\cal R} \sim H\delta q
/(\rho+P)$ for spatially flat hypersurfaces 
evolves during inflation to give significant
correction $\Delta \zeta^{(3rd)} \sim  H\delta \phi/\dot{\phi}$, which
finally meet the previous study of warm inflation.
{\bf The gauge invariance of the curvature perturbations and the
evolution is now clear in this formalism.}\footnote{Note that the
$\delta N$ perturbation defined by $H\delta \phi/\dot{\phi}$ is obviously
different from the conventional gauge-invariant definition of the
curvature perturbation. See also the appendix of this paper.}
The other is that evolution during inflation may create
{\bf another significant correction} $\Delta \zeta_N^{(rad)}$,
which has been overlooked in previous study but may be as large as
$\Delta \zeta_N^{(rad)} \simeq \Delta \zeta^{(3rd)}$ for $r_\Gamma
\simeq 1$. 
This term may leave significant signature in cosmological
perturbations when warm inflation passes from
strongly dissipating phase to weakly dissipating phase, which may help
reveal the mechanism of inflation.

\underline{$\Delta \zeta_N$ as the source of the cosmological perturbations}

Considering Eq. (\ref{normal-curv}) and Eq.(\ref{pert-phi}), we find for
the curvature perturbation created at the horizon exit;
\begin{eqnarray}
{\cal R}_{ini}&\sim& \frac{H\delta \phi}{\sqrt{\dot{\phi}^2} (1+r_\Gamma)}
\nonumber\\
&\sim&\frac{H^2}{T^2}\frac{r_\Gamma^{3/4}t_T^{1/2}}{(1+r_\Gamma)}
\nonumber\\
&\sim& \frac{r_\Gamma^{3/4}}{(1+r_\Gamma)r_T^{3/2}}
\sim r_\Gamma^{-{1/4}}r_T^{-{3/2}}.
\end{eqnarray}
If the CMB normalization is applied to $\Delta \zeta_N$ ({\bf not to
${\cal R}_{ini}$}), it is found that
\begin{equation}
\frac{r_\Gamma^{7/4}}{r_T^{3/2}(1+r_\Gamma)}\sim 10^{-5},
\end{equation}
where ${\cal R}_{ini}\sim r_\Gamma^{-{1/4}}r_T^{-{3/2}}<10^{-5}$ is assumed for the
domination by $\Delta \zeta_N$.
The condition for the successful creation of the curvature
perturbation in terms of $\Delta \zeta_N$ is thus given by
\begin{equation}
r_T^{-3/2}r_\Gamma^{7/4}\sim 10^{-5}.
\end{equation}

\underline{Inhomogeneous diffusion rate}

It would be useful to discuss another source for creating
inhomogeneities of the expansion rate during warm inflation.
Here we briefly show that inhomogeneities of the diffusion rate $\Gamma$,
which may 
be induced by light (moduli) fields, may cause inhomogeneities
of the expansion rate during warm inflation.
In fact, if dissipation occurs into light degrees of freedom and it
quickly thermalize into radiation, it is found that
\begin{equation}
\dot{\rho}_r + 4H \rho_r=\Gamma \dot{\phi}^2,
\end{equation}
and if the continuous dissipation keeps $\rho_r$ constant during warm
inflation, it leads to
\begin{equation}
\label{diss-pation}
4\rho_r \simeq 3 r_\Gamma \dot{\phi}^2.
\end{equation}
From the equation of motion, the slow-roll velocity of the inflaton
field is found to be given by
\begin{equation}
\dot{\phi}\simeq -\frac{V_\phi}{3H(1+r_\Gamma)}.
\end{equation}
For $r_\Gamma >1$, the fluctuation of the slow-roll velocity that arises 
from the inhomogeneous diffusion rate is thus given by
\begin{equation}
\frac{\delta \dot{\phi}}{\dot{\phi}}\simeq 
\frac{\delta \Gamma}{\Gamma},
\end{equation}
where the diffusion rate $\Gamma$ is diffent in diffent Hubble patches.
The inhomogeneities of $\Gamma$ may also create $\delta \rho_r$ in terms
of the dissipation given in Eq.(\ref{diss-pation}).
The inhomogeneous diffusion rate thus leads to inhomogeneities of the
expansion rate, which creates curvature perturbations given by 
\begin{equation}
\Delta_{(\delta \Gamma)} \zeta_N \sim \frac{\delta \Gamma}{\Gamma},
\end{equation}
where the magnitude of the perturbations is determined by the model.
A possible source of $\delta \Gamma$ is the perturbations related to
entropy (moduli) fields.
The diffusion rate may also depend on the temperature.
The temperature perturbation decays after horizon crossing,
however it may leave significant correction after time integration.

Introducing a light field $\sigma$, a specific example of the diffusion
rate can be given 
by\cite{Warm-decay}
\begin{equation}
\Gamma \sim C(\sigma) \frac{T^3}{\phi^2},
\end{equation}
which causes inhomogeneities of $\Gamma$ given by
\begin{eqnarray}
\frac{\delta \Gamma}{\Gamma}&\sim& \frac{C_\sigma}{C} \delta \sigma
+3\frac{\delta T}{T}
-2\frac{\delta \phi}{\phi}\nonumber\\
&\sim& \frac{C_\sigma}{C} \delta \sigma +3r_T^{-3/2}
-2r_\Gamma^{1/4}r_T^{1/2}\frac{H}{\phi}.
\end{eqnarray}
If is also possible to find significant non-gaussianity from $\delta \Gamma$,
expanding the moduli-dependent part as $\delta C  \simeq 
C_0 + C_2\frac{\sigma^2}{M_*^2}$, for example.
It is useful to specify the level of non-Gaussianity by the non-linear
parameter $f_{NL}$, which is usually 
defined by the Bardeen potential $\Phi$,
\begin{equation}
\Phi=\Phi_{Gaussian}+f_{NL}\Phi_{Gaussian}^2.
\end{equation}
Using the Bardeen potential, the curvature perturbation $\zeta$ is given
by
\begin{equation}
\Phi=\frac{3}{5}\zeta.
\end{equation}
Considering the expansion of $\delta N$ as
\begin{equation}
\zeta \simeq N_\phi \delta \phi + N_\sigma\delta\sigma
+\frac{1}{2} N_{\phi\phi} \delta \phi^2 
+\frac{1}{2} N_{\sigma\sigma} \delta \sigma^2 + ...,
\end{equation}
A useful simplification is\cite{Lyth_and_Rod_NG} 
\begin{equation}
f_{NL}\simeq \left(\frac{1}{1300}
\frac{N_{\sigma\sigma}(\delta \sigma)^2}{N_\phi^2 (\delta \phi)^2}\right)^3. 
\end{equation}
The non-linear parameter is estimated as
\begin{equation}
f_{NL}\sim 10^{21}C_2^3\left(\frac{\delta \sigma}{M_*}\right)^6,
\end{equation}
which may be as large as $f_{NL}\sim 10$.

\subsection{Decay into non-relativistic matter}

If the non-relativistic matter $\chi$ created by the dissipation of the
inflaton motion does not decay, the evolution equation
for the inflaton is given by
\begin{equation}
\ddot{\phi}+ (3H+\Gamma)\dot{\phi}+V^{eff}_\phi=0.
\end{equation}
Introducing entropy field ${\cal M}$ (moduli) in addition to the adiabatic
inflaton field $\phi$, the effective potential can be expressed as 
\begin{eqnarray}
V^{eff} &\equiv& V_0(\phi)+\rho_\chi\nonumber\\
&\simeq& V_0(\phi)+M_\chi(\phi,{\cal M})n_\chi.
\end{eqnarray}
Here $M_\chi$ and $n_\chi$ are the effective mass and the number
density of the non-relativistic matter $\chi$.
If the continuous dissipation keeps the energy density $\rho_\chi$
constant for a time period during inflation, and also the pressure is
given by using a constant $\omega$ as $p_\chi = \omega \rho_\chi$, the
energy density and the pressure are given by
\begin{eqnarray}
\rho &=& K + V^{eff}\nonumber\\
p&=& K-V^{eff} + (1+\omega) \rho_\chi.
\end{eqnarray}
Again, slow-roll approximation is useful in estimating the order of
magnitude of the physical quantities:
\begin{eqnarray}
\dot{\phi}&\simeq& -\frac{V_\phi}{3H(1+r_\Gamma)}\\
(1+\omega)\rho_\chi &\simeq& 2 r_\Gamma K\\
\rho+P &=& 2K + (1+\omega )\rho_\chi \simeq 2(1+r_\Gamma)K.
\end{eqnarray}
Therefore, on spatially flat hypersurfaces $\psi=0$,
comoving curvature perturbation created during warm inflaton
 is expressed as
\begin{eqnarray}
{\cal R}&=&  -H\frac{\delta q}{\rho+p}\nonumber\\
&\sim &  -H\frac{\delta q}{\dot{\phi}^2}\times (1+r_\Gamma)^{-1}.
\end{eqnarray}
Unlike warm inflaton, the source of the cosmological perturbations is
not thermal fluctuations but the conventional field perturbations
created during inflation.
The comoving density perturbation is given by
\begin{eqnarray}
\epsilon_m &\equiv& \delta \rho-3H\delta q\nonumber\\
&=&\delta K + \left(\delta V^{eff} - r_\Gamma^{-1} 
V^{eff}_\phi\delta \phi \right) -r_\Gamma^{-1} \ddot{\phi}\delta \phi.
\end{eqnarray}
It follows that
\begin{eqnarray}
-\dot{\psi} &\simeq&
H\frac{2 \left(\delta V^{eff} - r_\Gamma^{-1} 
V_\phi\delta \phi \right) 
- (1+\omega)\delta \rho_\chi}{2(1+r_\Gamma)K}\nonumber\\
&\sim&H\frac{(1-\omega)\delta \rho_\chi }{2(1+r_\Gamma)K}.
\end{eqnarray}
Using the above equations and the $\delta N$ formula, the evolution of
the curvature perturbation caused by the inhomogeneities of the
non-relativistic matter $\chi$ is given by
\begin{equation}
\dot{\zeta}_N\sim 
H\frac{(1-\omega)}{(1+\omega)}
\frac{r_\Gamma}{1+r_\Gamma}\frac{\delta \rho_\chi}{\rho_\chi}.
\end{equation}
Here $\delta \rho_\chi$ is sourced by $\delta \Gamma$, which arises due
to the inhomogeneities of the entropy fields.

\section{Conclusions and discussions}
In warm inflation with single-field inflaton, the effective potential of
      the inflaton field  
depends on the temperature $T$, which can be expressed as $V(\phi,T)$.
The dissipation rate $\Gamma$ may also depend on $T$ and $\phi$.
In this sense, the ``trajectory'' of inflation
is given by both $\phi$ and $T$.

The situation reminds us of multi-field inflation, in which
the dynamics of a inflaton field is determined by the dynamics of the
      other inflaton fields, and vice versa.
To compare the situations, first consider two-field inflation with a 
rapid-rolling inflaton $\phi_F$ that reaches its minimum during
inflation and a slow-rolling inflaton $\phi$ that determines the
number of e-foldings of the inflationary expansion.
Namely, on spatially flat hypersurfaces, the conventional definition of
the curvature perturbation is given by 
\begin{eqnarray}
{\cal R}_{multi}^{(ini)}=-H\frac{\delta q}{\rho+P}=
H\frac{\dot{\phi}_F\delta \phi_F+\dot{\phi} \delta \phi}
{\dot{\phi}_F^2+\dot{\phi}^2}
\simeq H\frac{\dot{\phi} \delta \phi}
{\dot{\phi}_F^2},
\end{eqnarray}
where $\delta \phi_F\simeq 0$ on the steep potential.
In this case the evolution during inflation is crucial, since 
at the ``bend'' of the inflation trajectory it gives
\begin{eqnarray}
\Delta{\cal R}_{multi}\simeq H\frac{\delta \phi}
{\dot{\phi}}\gg {\cal R}_{multi}^{(ini)},
\end{eqnarray}
which is consistent with the $\delta N$ formalism based on $\delta \phi$
and $\dot{\phi}$.
A similar situation appears in warm inflation, in which the curvature
perturbation is given by
\begin{eqnarray}
{\cal R}_{warm}^{(ini)}=-H\frac{\delta q}{\rho+P}=
H\frac{\dot{\phi}\delta \phi}
{\dot{\phi}^2+Ts}
\simeq H\frac{\dot{\phi} \delta \phi}
{Ts},
\end{eqnarray}
where the thermal perturbations disappear at large scales. 
Again, the evolution during inflation is crucial, since 
from a decaying component of warm inflation we find
\begin{eqnarray}
\Delta{\cal R}_{warm}\simeq H\frac{\delta \phi}
{\dot{\phi}}\gg {\cal R}_{warm}^{(ini)},
\end{eqnarray}
which is consistent with the previous study based on $\delta \phi$
and $\dot{\phi}$.

We also found that a significant correction may arise from the
perturbations of the radiation density, which raises significant 
correction at the intermediate region between strongly dissipating warm
inflation (SDWI) and weakly dissipating warm inflation(WDWI).

This paper considers warm inflation as an interesting application of 
 multi-field inflation. 
The $\delta N$ perturbation defined for a $\phi$-constant hypersurface
 may give the required curvature perturbation.
However, considering the conventional definition of the curvature
      perturbation given in Eq. (\ref{zeta-org}), which is manifestly
      gauge-invariant, there is a serious discrepancy between 
these two definitions.
The problem must be solved explicitly by considering gauge-invariant
      evolution of the curvature perturbation, which is
 defined by Eq. (\ref{zeta-org}).
Since a $\phi$-constant hypersurface  is identical
      to uniform density hypersurface at large scales,
the evolution at small scales must be very important in solving this
      problem.
Consideration of the decaying term in the evolution of the curvature
      perturbation is a new idea, which is necessary in solving this
      problem.
A solution to this problem is one of the main result obtained in this
paper.

Significant corrections to the curvature perturbation are raised
from the small-scale perturbations of the radiation and from the
decaying $\delta 
\phi$, which are found to dominate the
 curvature perturbation.
The one from the decaying $\delta \phi$ reproduces the $\delta N$
perturbation for $\phi$-constant hypersurfaces, and the other from the
radiation creates a new contribution, which can nearly dominate the
curvature perturbation for $r_\Gamma \simeq 1$.

It is also shown that significant non-gaussianity may be created from
the inhomogeneities of the dissipation rate, which appears independent
of the 
conventional curvature perturbations.

The short-scale corrections that we have considered in this
paper may be dubbed the non-equilibrium corrections.
Other kinds of corrections have already been discussed for
supercooled inflation in
Ref.\cite{Modulated-matsuda}, and these may be significant in string
cosmological models.

\section{Acknowledgment}
We wish to thank K.Shima for encouragement, and our colleagues at
Tokyo University for their kind hospitality.

\begin{appendix}
\section{``Local'' and ``global'' definitions of the $\delta N$ formula}
In this paper we considered the $\delta N$ formula defined for the
uniform density hypersurfaces at a time-slice and calculated the
evolution of $\delta N$.
In this appendix, the $\delta N$ formula defined in this way is denoted
by $\delta N_{local}$.
It is also possible to define $\delta N$ using the number of e-foldings 
$N_e$, which is defined using the interval between the horizon exit at
$t=t_{ini}$ and the end of inflation at $t=t_e$. 
We denote the latter definition by $\delta N_{global}$.
Obviously, it is impossible to calculate the ``evolution'' (not the
scale-dependence)  of $\delta N_{global}$.\footnote{Note also that the
curvature perturbations defined at the horizon exit (${\cal R}_{ini}$)
has the trivial $k$-dependence, which must be distinguished from the
``evolution'' defined in this paper.} 

$\delta N_{local}$ is defined so that it is related to the curvature
perturbations at the time-slice.
If one wants to understand the evolution of the curvature perturbations
in terms of the $\delta N$ formalism, one should consider $\delta
N_{local}$ for the definition of the $\delta N$ formula.

One may find a discrepancy between $\delta N_{local}$ and 
$\delta N_{global}$, which must be solved by considering the evolution of
$\delta N_{local}$.
A simple example of the inflaton trajectory in typical multi-field
inflation is 
shown in Figure 1.
Here we consider two-field inflation with a separable potential.
The $\phi_2$-potential is steep and it rolls much faster than $\phi_1$.
Following the standard definition of the curvature perturbations, we find
that the curvature perturbations at a time-slice is given by
\begin{eqnarray}
{\cal R}_{ini}=-H\frac{\delta q}{\rho+P}=
H\frac{\dot{\phi}_1\delta \phi_1+\dot{\phi}_2 \delta \phi_2}
{\dot{\phi}_1^2+\dot{\phi}_2^2}
= H\frac{\delta \phi_\sigma}
{\dot{\phi}_\sigma},
\end{eqnarray}
where the adiabatic inflaton is defined by 
$\dot{\phi}_\sigma^2 \equiv \dot{\phi}_1^2 +\dot{\phi}_2^2$.
However, one may find from the trajectory (See figure 1) that the number
of e-foldings can be determined exclusively by $\phi_1$, and there
may not be a $\delta \phi_2$-dependence in $\delta
N_{global}$.\footnote{More examples can be found in the first paper in
\cite{End-multi-mat}, 
in which the fluctuations at the end-boundary are also 
taken into account for the calculation of $\delta N_{global}$.}
On the other hand, if one defines the curvature perturbations at a time
slice by ${\cal R}\sim H\delta \phi_1/\dot{\phi}_1$, one immediately
finds that the definition is not manifestly gauge-invariant.
In this case the evolution during inflation is crucial, since 
at a ``bend'' of the inflation trajectory the evolution is given by
\begin{eqnarray}
\Delta{\cal R}\simeq H\frac{\delta \phi_1}
{\dot{\phi}_1}\gg {\cal R}_{ini}.
\end{eqnarray}
One might claim that the argument related to the evolution at the
``bend'' of the trajectory gives a trivial argument for the curvature
perturbations, claiming that it gives a trivial result in view of
$\delta N_{golbal}$. 
However, it is not obvious if the intuitive argument
based on $\delta N_{global}$ always gives the correct result.
Therefore, the source that explains the gap in
these definitions must be identified from the calculation of the
evolution of $\delta N_{local}$.

\begin{figure}[h]
 \begin{center}
\begin{picture}(200,200)(100,350)
\resizebox{15cm}{!}{\includegraphics{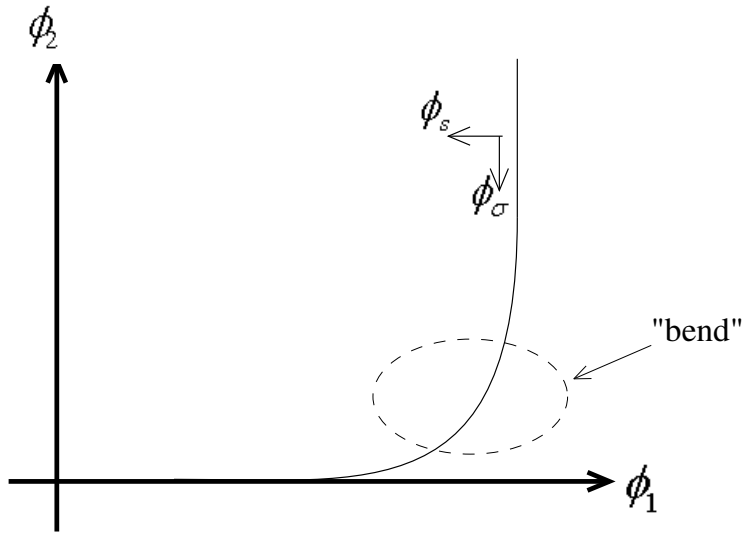}} 
\end{picture}
\label{fig:multi-basic}
 \caption{The curved line represents a trajectory in two-field inflation.
$\phi_2$ rolls faster than $\phi_1$ and is significant at the beginning.
However, soon $\phi_2$ reaches its minimum and the
  dynamics of $\phi_2$ is negligible near the end of inflation. 
Note also that it is possible to consider a $\phi_2$ field that is heavy
  and oscillating during inflation.
In this case, $\phi_2$ generates ``many bends'' in the trajectory. 
See also the first paper in Ref.\cite{End-multi-mat} for more details of
  the global calculation of $\delta N$.} 
 \end{center}
\end{figure}
\end{appendix}

Applying the same argument to warm inflation, it is easy to find that 
the usual definition of
the curvature perturbations in warm inflationary models are based on
$\delta N_{global}$, while in 
this paper we calculated the evolution of $\delta N_{local}$ to find
that the calculation based on the intuitive argument of $\delta
N_{global}$
is consistent with the explicitly gauge-invariant
definition of the curvature perturbations at a time-slice.
In our calculation we also introduced another inflaton field to 
show what happens at the bend of the trajectory in warm inflation.
Note that we are not considering a counter-example that may ruin the
usual argument based on $\delta N_{global}$.

The end-boundary of warm inflation is usually defined by the field value
at a critical point where slow-roll condition is violated.
However, more generically the end-boundary can be defined by a critical
temperature, as will be discussed in our forthcoming
paper\cite{remote-inflation}.

\end{document}